# Multiple model approach and experimental validation of a residential air to air heat pump.

François GARDE *Research Engineer, Electricité de France / University of Reunion Island.* \*
Harry BOYER, *Assistant Professor, University of Reunion Island*\*
Florence PIGNOLET, *Researcher, University of Reunion Island*\*
Franck LUCAS, *Researcher, University of Reunion Island.*\*
Jean BRAU, *Professor, INSA de Lyon, CETHIL/ Thermique du Bâtiment†*

**Abstract :**
    The beginning of this work is the achievement of a design tool, which is a multiple model software called « CODYRUN », suitable for professionnals and usable by researchers. The original aspect of this software is that the designer has at his disposal a wide panel of choices between different heat transfer models More precisely, it consists in a multizone software integrating both natural ventilation and moisture tranfers . This software is developed on PC micro computer and gets advantage of the Microsoft WINDOWS front-end.
    Most of time, HVAC systems and specially domestic air conditioners, are taken into account in a very simplified way, or in a elaborated one. On one side,they are just supposed to supply the demand of cooling loads with an ideal control loop (no delay between the sollicitations and the time response of the system), The available outputs are initially the hourly cooling and heating consumptions without integrating the real caracteristics of the HVAC system This paper is also following the same multiple model approach than for the building modelling by defining different modelling levels for the air conditionning systems, from a very simplified one to a detailed one.
    An experimental validation is achieved in order to compare the sensitivity of each defined model and to point out the interaction between the thermal behaviour of the envelop and the electrical system consumption. For validation purposes, we will describe the data acquisition system. and the used real size test cell located in the University of Reunion island, Indian Ocean.

## 1. INTRODUCTION

   The main problem of Demand Side Management in islands such as Reunion island is that the means of electrical production are restricted. The demand of power supply is not so important to justify a nuclear power plant, and the means of production are more expensive than the nuclear. In that case, cooling of buildings and interaction between the envelop and the cooling system is a growing problem because air conditioning systems consume a large amount of energy and generate investments in power supply. In addition, building envelops suffer from a bad thermal conception that generates high cooling loads. The large population increase in the tropical islands, the rise in the living standards and the decreasing costs of air-conditioning units constitute a real energetic problem.
   The modeling and simulation of HVAC systems , and the interaction with the building is  also necessary for the evaluation of means for reducing energy use.
   We focus in our studies on the modelling of a residential air-to-air split-system heat-pump, which is the more widespread kind of air conditioning system in tropical islands. The problem in heat-pump modelling is the level accuracy of the model. Most of time, HVAC systems and specially domestic air conditioners, are taken into account in a very simplified way, or in a elaborated one. The question is : Untill what kind of degree of complexity the heat-pump can be modelled for a further integration in a thermal and airflow simulation software. According to Irving [15], there is an important number of simulations softwares of varying degrees of complexity available in the industry. These programs range from very simple steady-state heat or cooling loads calculation running on programmable calculators, to very detailed simulation programs which require large amounts of computer power. It is obvious that no single program can satisfy the many needs of the engineer at every stage of his design.

Another point to take into account is the time step of the calculation programs. It is often of one hour, what is too long to take into account the HVAC systems control, because their time constant are shorter than the building ones. It is apparent that optimal cooling control strategies require a dynamic model for predicting the performance characteristics of the overall system [10]. At last, the cooling coil of the indoor unit of the split-system is one of the key component in air conditionning system. The main difficulties are located in wet and/or transient regimes and the performance of the whole system (and thus the accuracy of the model) depends on a good model of the cooling coil.

## 2. MODELLING REVIEW

Concerning heat-pump simulation, it appears that the different models can be broadly classified as steady and dynamic state simulations. Weslby [25] reviewed with particular emphasis on their bases and end-uses various mathematical models on mechanical vapour-compression heat-pumps since from 1975 to 1988.

It appears that detailed steady state models are often used to study variations in system or component configuration in order to identify optimal values of parameters that dictate the performance of the system. They are very helpful to optimize the design of air conditioning equipments. The component approach is often used by researchers because of the strong coupling between them. [2]. In a general way, authors have developed a numerical model to determine the steady-state performance of heat-pumps for specified source and sink conditions and specified components. The model is based on simple models of the components of the heat pump. Considering the coupling with a thermal simulation software, Loveday [16] has developed a heat-pump simulation programme in order to relate the hourly thermal building loads to the corresponding electricity inputs to the heat-pump, which is a function of its COP. The hourly heating energy to be supplied to the building is obtained from ESP [9]. The heat-pump model is a steady-state model which relates heat-pump COP to external ambient temperature, and accounts for motor/compressor inefficiencies, thermal/mechanical losses of the heat transport system, and defrost performance.

These steady-state models are most of time too detailed for the use we need. Other investigators have worked on steady-state models based on experimental data or manufactured data [1], [12], [24]. From the knowing of differents set points and different experimental conditions, a model based on polynomial laws relating the cooling capacity or the power consumptions to the outdoor and indoor temperature can be elaborated.

Transient perturbations and dynamic regimes could be simulated to analyse the start-up process, defrost-cycle and rapidly varying operating conditions. The aim of these models is to study the variation of working fluid mass and thermodynamic state distributions within the system from known initial conditions, when the system operating conditions are changed. Some reseachers have developed rigourous dynamic models [8], [19], [17], [23]. These models have ranged from lumped parameters [19], [23] to mathematical models based on a multi-node or distributed approach [17]. These models tend to be complex, require large computers to use, and do not readily yield physical insight into the variables affecting system performance. However, they do allow the user to track detailed information (temperatures, pressures, etc.) in the system during the transient start-up process.

Other investigators [18] have suggested a two-time constant model to capture the physical phenomena responsible for strat-up losses. One time constant would capture capacity delay due to the mass of the heat exchanger. The second time constant would capture a very high, but rapidly decaying, initial capacity that is produced by the time lag for the refrigerant to be pumped from the evaporator to the rest of the system. This model should provide a better fit to the experimental data during the start-up process.

Other investigators have hypothized that during start-up, the system capacity could be modeled as a first-order (single-time constant) system [7], [13], [20], [21], [22]. Experimental data from Murphy [20] showed good agreement between a single time constant model and data from a heat-pump. We also verify that hypothesis by ourselves and have found time constant of 2 minutes. Murphy found time constants for cooling mode operations ranging from 0.32 to 0.47 minutes for a heat-pump and an air-conditionner respectively. O'Neal [22] found time constant of more than 2 minutes for a heat-pump operating in cooling mode. Another important assumption made by Glosdsmith [13] and O'Neal [21] was the constant power upon star-up. This assumption seems to be reasonable. While there is an initial surge in power during the few seconds after start-up, the power is relatively constant during start-up. We also verify that point by our own experiments and didn't notice a high surge in power after start-up of the split-system. Finally, because the aim of our study was to take into account the exponential increasing of cooling capacity and not specially the mecanisms affecting cycling, the choice was made to use a single-time constant model.

Considering the coupling effects of cooling and deshumidifying on a cooling coil, several authors have proposed different models of coil in steady-state and/or transient conditions. Hirsch [14] considers that the moisture condensation on cooling coils is simulates by characterizing the coils by their bypass factors and solving the bypass relation with the system moisture balance in steady-state conditions. Xin Ding [26] has worked on different models of cooling coil in steady-state and transient conditions, and in dry or wet regimes. These models are based on the NTU method and on the determination of the effectiveness of the coil in wet and dry regimes. The dynamic model of the coil is supposed to be a first order model.

### 3. CODYRUN AND THE INTEGRATED EQUATIONS FOR HVAC SYSTEMS

*3.1 An overview of* CODYRUN *software :*

CODYRUN is a thermal multizone software integrating both natural ventilation and moisture transfers. Its main characteristics is to be a multiple model structure, allowing the choice between a wide range of models of heat transfer and meteorological data reconstitution. More information can be obtained about the software in [3] concerning multiple model aspect, building description in [5], thermal model constitution in [4] and preliminary software validation in [11].

Concerning the calculation, the main parts are the airflow model (pressure model taking into account wind, thermal buoyancy and large openings), the thermal model and the moisture transfer model. For the study presented in this paper, the two last models has been modified by keeping the same multiple model concept, allowing finally different levels of modelisation for the air cooling system.

*3.2 Quick thermal and moisture transfer models constitution :*

By considering usual assumptions as mono-dimensional heat conduction in walls, well mixed air volumes, and linearized superficial exchanges, nodal analysis (or lumped capacities analysis) leads to an electrical network. To simplify our discussion, we'll suppose that the heat conduction is treated with the help of a model constituted of a thermal resistance and 2 capacitors, said "*R2C*" model (leading to no internal nodes in walls). Then, the thermal model of the building is obtained as a set of equations of type from 1 to 4, traducing thermal balance of inside (*Tsi*) and outside nodes (*Tse*) according to boundary conditions, thermo-convective balance equation of dry-bulb air nodes (*Tai*) and radiative balance equation of the inside mean radiant temperature nodes (*Trm*).

$$C_{si} \frac{dT_{si}}{dt} = h_{ci}(T_{ai}-T_{si}) + h_{ri}(T_{rm}-T_{si}) + K(T_{se}-T_{si}) + \varphi_{swi} \qquad (1)$$

$$C_{se} \frac{dT_{se}}{dt} = h_{ce}(T_{ae}-T_{se}) + h_{re}(T_{ae}-T_{se}) + K(T_{si}-T_{se}) + \varphi_{swe} \qquad (2)$$

$$C_{ai} \frac{dT_{ai}}{dt} = \sum_{j=1}^{Nw} h_{ci}(T_{ai}-T_{si(j)}) + c\dot{Q}(T_{ae}-T_{ai}) + \dot{Q}_{sens} \qquad (3)$$

$$0 = \sum_{j=1}^{Nw} h_{ri} A_j (T_{si}(j)-T_{rm}) \qquad (4)$$

Calculating separately moisture transfers (induced by air motion between indoor zones and outdoor, but without latent storage in walls and room furniture*),* for a zone of specific humidity $r_s$ , moisture balance leads to a linear system of equations, each being as

$$C \frac{dr_s}{dt} = \dot{m}_{in} l_v r_{s\,in} - \dot{m}_{out} l_v r_s + \dot{Q}_{lat} \qquad (5)$$

$\dot{Q}_{sens}$ and $\dot{Q}_{lat}$ are respectively is the sensible and latent capacities injected in the zone, even by internal loads (lighting, occupants) or HVAC systems.

*3.3 Integration of air cooling system :*

At each time step, depending on indoor and outdoor conditions, in case of air cooling, the HVAC model has to calculate $\dot{Q}_{sens}$ and $\dot{Q}_{lat}$ for the considered zones, values to be injected in equations *(3)* and *(5)*.

**4. MODELS DEVELOPMENT**

*4.1 Model n°0*

This model is just supposed to supply the demand of cooling loads with an ideal control loop (no delay between the sollicitations and the time response of the system), The available outputs are initially the hourly cooling consumptions without integrating the real caracteristics of the HVAC system. We assume that sensible latent cooling rates are not dependant. The real physical phenomenon are different beacuse both cooling and deshumidifying occur at the same time.

The HVAC system must reach a temperature set point and a humidity set point. It allows the user to determine the hourly sensible and coolings loads of the building, but does not depend on some specific characteristics of the system.

In order to arrive at electric consumption, the system is modelled by its cooling Coefficient of Performance (COP), which is the ratio between the cooling rate and the electric power. The COP is taken constant in our case.

The COP is also a first comparison quality criteria for heat pump systems . It allows to detemine the electric consumption of the system all along the simulation period.

*4.2 Model n °1:*

This model is more detailled than the first one. The time step was reduced to one minute in order to take into account the on-off cycling and the control of the system.

*Steady -state conditions :*

The capacity rate is the nominal capacity of the system operating in steaty state and standard conditions. (27°C dry bulb temperature, 19°C wet bulb temperature, 35°C outdoor temperature). The system is defined in steady state conditions by its SHF (sensible heat factor) which is the ratio between the sensible cooling rate and the total cooling rate. The SHF gives also the sensible and latent capacities which will be integrated at each time step in the solving equations of CODYRUN. These capacities are held constant each time the split is on.

*Dynamic conditions :*

As seen in the modelling review part, the system capacity is modeled as a first order model, in order to take into account the start-up of the split-system [21], [22].
Figure 1 shows the idealized capacity during start-up and shutdown of a residential air-conditioner.

The instantaneous cooling capacity, $\dot{Q}cyc$ is given by :

$$\dot{Q}cyc = \dot{Q}ss(1 - e^{-t/\tau}) \qquad (6)$$

where  $\dot{Q}cyc$  : steady state capacity
        $\tau$  : time constant for the system
        t  : time after start-up

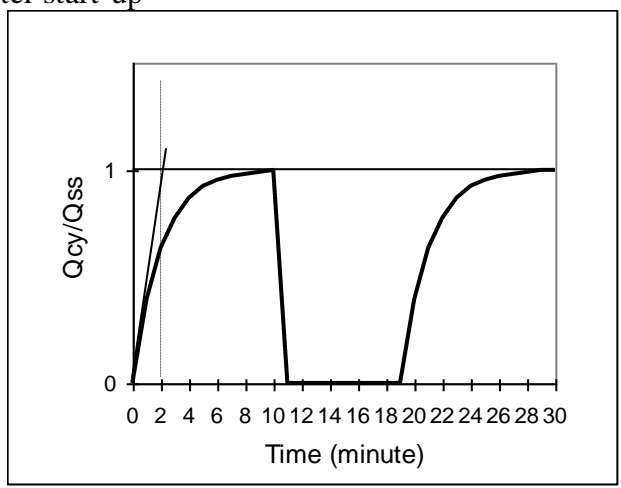

*Fig 1 : Dynamic behaviour during start-up condition.*

*System control :*

An on/off controller with a dead zone has been developed. This kind of controller are always used for small sized residential unit. Some experiments allowed us to find the correct dead zone of the controller, which was found equal to 1°C.

*Electric power :*

The electric power of the system is assumed to be constant, even during start-up conditions [13], [21], [22]. The user enters the cooling COP in nominal conditions in order to calculate the electric consumption of the system.

*4.3 Model n°2 :*

The performances of an air-to-air heat-pump (total cooling capacity, sensible and latent capacities, electric power) are strongly influenced by parameters such as the indoor dry air temperature, the indoor relative humidity and the outdoor air temperature. In model 1, the performances are supposed to be constant, but figure 1 points out that they depend on external parameters. The aims of these models are to take into account these effects in steady-state conditions.

*Steady-state conditions :*

The method is based upon manufacturer data given in Table 1. These data were mesured during steady-state conditions. The indoor dry-bulb temperature was held constant. Only parameters such as wet bulb temperature and outdoor temperature were changing. We simply add to table 1 the values of the indoor air specific enthalpy.

| *CARRIER Indoor unit : 42 HWA - Outdoor unit : 38 SDF012* | | | | | | |
|---|---|---|---|---|---|---|
| Indoor dry bulb temperature : 27°C | | | | | | |
| | | By Pass factor = 0.04 | | Air flow rate : 110 l/s | | |
| | Wet bulb temp (°C) | 16 | 18 | 19 | 20 | 22 |
| Outdoor temp (°C) | Indoor air specific enthalpy (kJ/kg) | 45 | 55 | 58 | 61.2 | 65 |
| 21 | Total capacity | 3.46 | 3.65 | 3.72 | 3.78 | 3.91 |
| | Sensible capacity | 3.26 | 2.88 | 2.67 | 2.45 | 2.04 |
| | Power (kW) | 0.99 | 1.00 | 1.00 | 1.01 | 1.02 |
| 25 | Total capacity | 3.34 | 3.54 | 3.64 | 3.72 | 3.84 |
| | Sensible capacity | 3.20 | 2.84 | 2.65 | 2.44 | 2.03 |
| | Power (kW) | 1.06 | 1.07 | 1.07 | 1.08 | 1.09 |
| 30 | Total capacity | 3.16 | 3.37 | 3.48 | 3.59 | 3.79 |
| | Sensible capacity | 3.10 | 2.78 | 2.59 | 2.41 | 2.03 |
| | Power (kW) | 1.14 | 1.15 | 1.16 | 1.17 | 1.19 |
| 35 | Total capacity | 2.95 | 3.18 | ***3.30*** | 3.40 | 3.63 |
| | Sensible capacity | 2.95 | 2.70 | ***2.52*** | 2.34 | 1.98 |
| | Power (kW)kW | 1.23 | 1.25 | ***1.25*** | 1.27 | 1.28 |
| 40 | Total capacity | 2.67 | 2.77 | 2.95 | 3.18 | 3.44 |
| | Sensible capacity | 2.67 | 2.52 | 2.38 | 2.25 | 1.91 |
| | Power (kW) | 1.33 | 1.34 | 1.35 | 1.36 | 1.38 |
| 45 | Total capacity | 2.49 | 2.56 | 2.69 | 2.93 | 3.3 |
| | Sensible capacity | 2.49 | 2.43 | 2.28 | 2.16 | 1.86 |
| | Power (kW) | 1.40 | 1.40 | 1.41 | 1.42 | 1.45 |

*Table 1 : Manufacturer data for a residential air-to-air split-system heat pump- Source : CARRIER Corp. Professionnal catalogue for residential units, 1995-1996. (The nominal values are noticed in italic and heavy fonts)*

We assume that for a fixed outdoor temperature, the sensible and total capacities are following linear laws. These equations can be written as following :

$$\dot{Q}tot(27°C) = a_0 + a_1 \cdot h_{ent} \qquad (7)$$

$$\dot{Q}sens(27°C) = b_0 + b_1 \cdot h_{ent} \qquad (8)$$

where :  $h_{ent}$ is the air entering specific enthalpy at the evaporator.
$a_0$, $a_1$, $b_0$, $b_1$ are the correlation coefficients of $\dot{Q}tot$ and $\dot{Q}sens$ respectively.

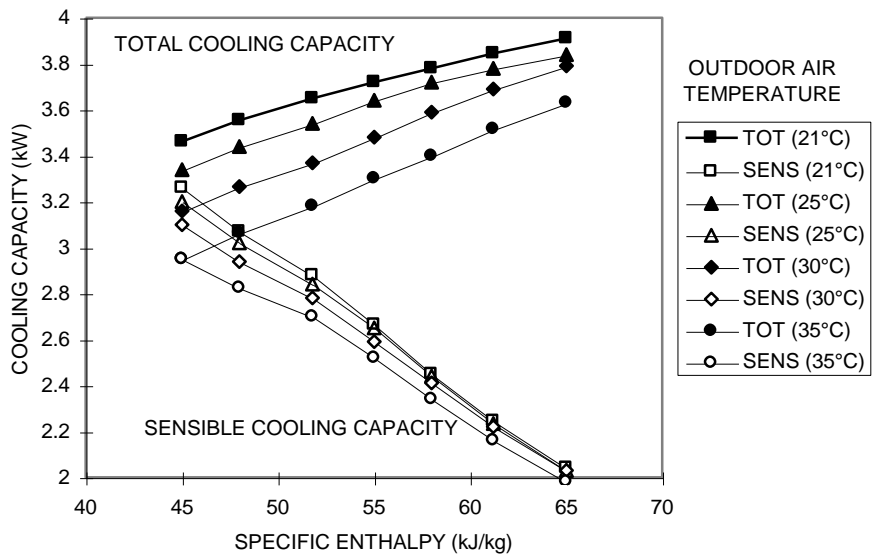

*Fig 2 : Evolution of total cooling capacity and sensible cooling capacity according to the indoor air specific enthalpy and for different outdoor air temperatures. Indoor air temperature = 27°C*

These coefficients depend on the outdoor air temperature. We determined the values of these coefficients for outdoor temperatures of 21°C, 25°C, 30°C and 35°C. We didn't take into account the values of 40°C et 45°C because on the one hand we are outside the operating range of the split-system, and on the other hand the evolution of the capacities tend to becom non linear.

These coefficients are also approximated by a linear law, so they can be expressed as :

$$a_0 = c_1 + c_2 \cdot Text \qquad (9)$$
$$a_1 = c_3 + c_4 \cdot Text \qquad (10)$$
$$b_0 = c_5 + c_6 \cdot Text \qquad (11)$$
$$b_1 = c_7 + c_8 \cdot Text \qquad (12)$$

where Text is the outdoor air temperature.

$c_1, c_2, c_3, c_4, c_5, c_6, c_7, c_8$ are the calculated correlation coefficients coming from the linear equations (9), (10), (11), (12).

Thanks to these correlations, we can expressed the total capacity and the sensible capacity by the following equations for an indoor air temperature of 27°C :

$$\dot{Q}tot(27°C) = c_1 + c_2 . Text + (c_3 + c_4 . Text).h_{ent}$$
$$\dot{Q}sens(27°C) = c_5 + c_6 . Text + (c_7 + c_8 . Text).h_{ent}$$

The manufacturer data have allowed us to estimate the cooling capacities, both total and sensible, as function of the entering specific enthalpy and the outdoor air temperature for the set temperature of 27°C. Now, we have to determine the cooling capacities at any operating point. Thus, we introduce the coil bypass factor concept [6], [14].

The coil bypass factor (BF) model characterizes the air exiting the coil as being composed of two major streams : the air which has not been influenced by the coil and the air which leaves at the coil surface temperature (apparatus dew point). The coil bypass factor is the fraction of air which is unaffected by the coil. Thus, we have relations for the exit dry-bulb temperature and humidity ratio in terms of entering conditions and the coil bypass factor.

$$BF = \frac{T_{out} - T_{adp}}{T_{ent} - T_{adp}} = \frac{h_{out} - h_{adp}}{h_{ent} - h_{adp}} = \frac{w_{out} - w_{adp}}{w_{ent} - w_{adp}} \qquad (15)$$

with $T_{adp}$ : apparatus dew point temperature or temperature of the coil surface
$h_{adp}$ : specific enthalpy on the coil surface.

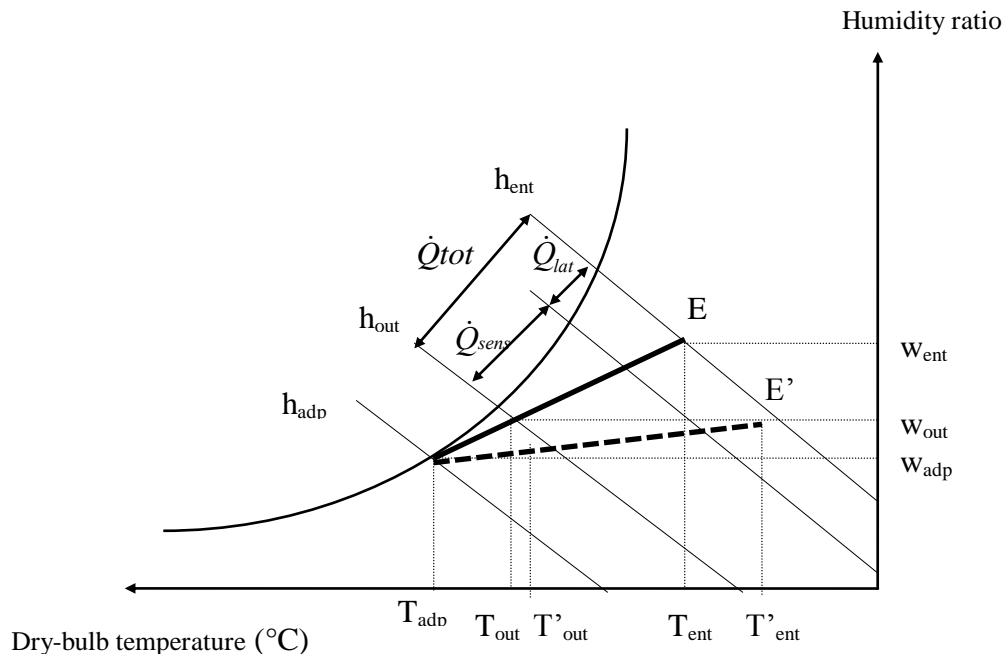

*Fig 3 : Cooling coil performance and determination of cooling capacities*

The coil bypass factor is a function of both physical and operation parameters of the coil : these parameters are the coil exchange surface *A* (when *A* is rising, *BF* is decreasing), and the coil air velocity *v* (when *v* is creasing, *BF* is creasing). We assume in our model that these parameters are constant (the air velocity is supposed to stay in the high position), so the bypass factor is held constant too.

Figure 3 shows an illustration of the problem. We made one important assumption in developping this model :

The total cooling capacity rate was constant for two operating points E and E' having the same entering specific enthalpy. That means that :

$$\dot{Q}tot(E) = \dot{Q}tot(E') = \dot{m} \cdot (h_{out} - h_{ent}) \qquad (16)$$

where $\dot{m}$ is the mass flow rate of the air through the coil (kg.s$^{-1}$)

The apparatus dew point is by consequence the same for the two entering points :

$$h_{adp}(E) = h_{adp}(E')$$

The sensible and latent capacities are changing between the two points, but the total capacity remains constant. Considering the point E', the equations (13) and (14) give the value for the same specific enthalpy of the total capacity and the sensible capacity for a drybulb temperature of 27°C.

In that case :

$$\dot{Q}sens = \dot{m} \cdot (h_{out} - h_B) = \dot{m} \cdot c_{pm} \cdot (T_{out} - T_E) = \dot{m} \cdot c_{pm}(1-BF)(T_{adp} - T_E)$$

$$\dot{Q}sens' = \dot{m} \cdot (h'_{out} - h_{E'}) = \dot{m} \cdot c_{pm} \cdot (T'_{out} - T_{E'}) = \dot{m} \cdot c_{pm}(1-BF)(T_{adp} - T_{E'})$$

$$\dot{Q}sens' = \dot{m} \cdot c_{pm}(1-BF)(T_{adp} - T_E) + \dot{m} \cdot c_{pm}(1-BF)(T_E - T_{E'})$$

$$\text{then} \quad \dot{Q}sens' = \dot{Q}sens + \frac{\dot{V}}{v} c_{pm}(1-BF)(27 - T_{E'}) \qquad (17)$$

where
$\dot{Q}sens'$ is the sensible capacity of the heat-pump for the E' entering point.
$\dot{Q}sens$ is the sensible capacity of the heat-pump for the E entering point ($T_E = 27°C$) given by the expression (14).
$\dot{V}$ is the air flow rate. It is held constant and equal to 0.136 m3.s$^{-1}$
$v$ is the specific volume of the air (m3.kg$^{-1}$).
$c_{pm}$ is the specific heat of the mixing of dry air and water vapour. It is assumed to be constant : $c_{pm}$=1.02kJ.kg-1.

We can determine the latent loads thanks to the expressions (16) and (17):

$$\dot{Q}tot(E) = \dot{Q}tot(E')$$
$$\dot{Q}sens' = \dot{Q}sens + \frac{\dot{V}}{v} c_{pm}(1-BF)(27 - T_{E'})$$
$$\dot{Q}lat' = \dot{Q}tot - \dot{Q}sens' \qquad (18)$$

Then, the values of $\dot{Q}sens'$ and $\dot{Q}lat'$ are integrated at each time step in equations (3) and (5).

*Dynamic state conditions :*
   The taking into account of the dynamic effects upon start-up is the same than in model n°1 (single time constant).

*System control :*
   The system control is an on/off controller with a dead zone of 1°C.

*Electric power :*
   We assumed that the electric power is a linear dependant fonction of the outdoor temperature. Figure 4 shows a good validation of this assumption. We assume that there is no dynamic effects during the few time steps after start-up.

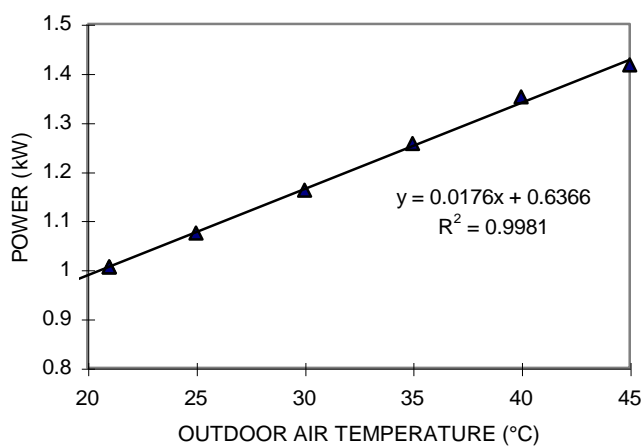

*Fig 4 : Electric power of the split-system*

### 5. EXPERIMENTAL SET-UP AND TEST PROCEDURE

A single-speed air-to air heat pump was instrumented and tested in a real size test cell located at the University of Reunion Island. The experiments were conducted in natural climatic conditions.

*5.1 Test split-system :*
   The test unit is a R22 air-to-air split-system residential heat pump that is commercially available. This split-system is operating only in cooling mode, and has a 3.3kW nominal cooling capacity. Both indoor and outdoor heat exchangers are of tube-and-plate-fin construction. The compressor is a hermetic rotative compressor and the expansion device is a capillary tube.
   Monitoring included the temperatures of the refrigerant fluid from both sides of each components (indoor and outdoor units, compressor and expansion device), the R22 volume flow rate by means of a liter-meter, high and low pressures, power and energy supply (active, apparant and reactive) by means of an electronic watt-meter. Even the temperatures of the return air and air outlet of the indoor unit the mass flow rate of the water coming from the indoor unit are mesured. Thus we can determine the total cooling capacity thanks to the thermodynamic cycle of the R22, and both sensible and latent capacities with the measurements of return and exit air of the indoor unit, and the measured mass flow rate of the water respectively.

*5.2 Test cell*

The size of the test-cell is 3.0 x 3.0 x 2.30m ( see fig. 5).
The test cell is made of sandwich panels consisting of two 7 mm layers of ciment-fiber boards with 6 cm of polyurethane foam between them. A layer of 5 cm of styrofoam is put between the floor panel and the concrete. On the roof, there is an extra sandwich panel, made of aluminium sheets and polyurethane, about 5 cm thick. The inner floor is made of 5cm concrete paving stones .

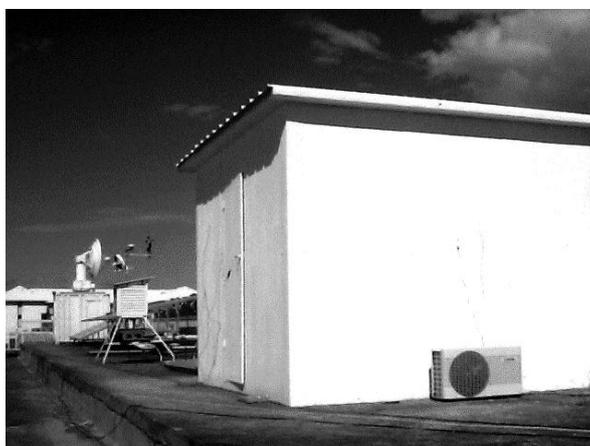

*Fig 5 : Vue of the Test cell and the outdoor unit of the split-system.*

Indoor air temperatures are mesured by type T (copper-constantan) thermocouples shielded against radiation, at three different levels (0.30m, 1.20m, 1.80m). External and internal temperature are mesured with type T thermocouple. The sensors are glued to the walls and painted with the same colours as the walls (i.e for the roof).

The indoor relative humidity is also recorded thanks to a temperatue and relative humidity probe.

*5.3 Weather data acquisition :*

A weather station was used close to the test cell (see fig.5 ) to measure the outdoor air temperature, the outdoor relative humidity, wind speed and direction and the global and diffuse horizontal solar radiations. Thanks to these data, the meteorological file of will be able to be monitored in vue to compare measurements and modelisation.

*5.4 Data acquisition system and computer processing.*

All the data are mesured and collected thanks to dataloggers that are controlled and programmed by a PC computer via RS232 links. The time step of the data acquisition can be modified function of our needs. For instance, we use a time step of one hour for the validation of the test cell without HVAC system, but the time step is reduced to one minute when the expriments are conducted with active systems.

*5.5 Experimental procedure*

The experiments were conducted at different steps. The first step consisted in a comparison between the simulation and the mesurements on the test cell without air-conditioning system in order to optimize the modelisation of the envelop. Figure 6 shows a passive period of two weeks. The measured and simulated air temperatures are very similar, with deviations smaller than 1°C.

In a second time, a lot of experiments were conducted in order to get accurate informations about the behaviour of the split-system unit (determination of the time constant for the model n°1 and 2, values of the dead zone band of the controller, etc...). Thanks to these preliminary tests we determined the time constant of the unit, which is equal to 2 minutes, and the dead zone band which is in a range of $\pm$ 0.5°C from the set temperature.

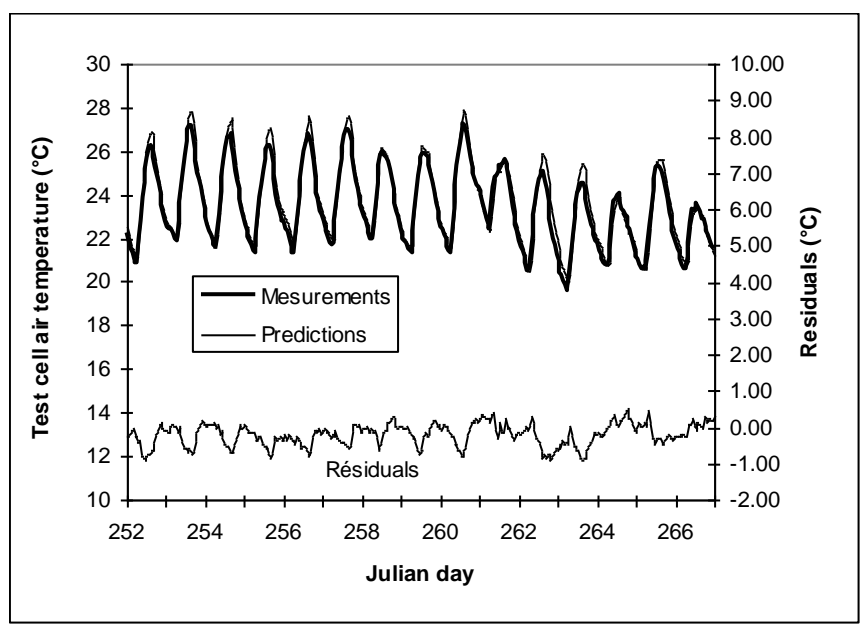

*Fig 6 : Comparison between mesurements and predictions - Passive period*

On the third time, experiments were conducted with the split-system in the operating mode for different indoor drybulb set temperatures. The time step of the experiments was taken to 1 minute. The test periods lasted generally for seven days . The first results are detailed in the following part

## 6. RESULTS

The temperature of the heat-pump was set to 23°C during the whole period of the measurements. The air renewal was neglected. There were no internal loads inside the test cell.

Figures 7 to 10 present measured data and simulation results of models 1 and 2 for a time step of one minute. The presented outputs in figures 7, 8, 9,10 were the indoor air temperature, the electric power, the sensible coling capacity and the total cooling capacity respectively The selected day belongs to a test period of seven days. (Oct. 20-27, 1996). We have pulled out the best day of the test period.

The model n°0 requires a time step of one hour. In view to compare the model n°0 to the simulations of models n°1 and 2 and to measurements, the results for a time step of one minute were averaged to one hour. Thus, Figures 11 to 13 presents the measured and simulation results for a time step of one hour. The presented outputs are the average indoor air temperature, the total cooling capacity and the electric power.

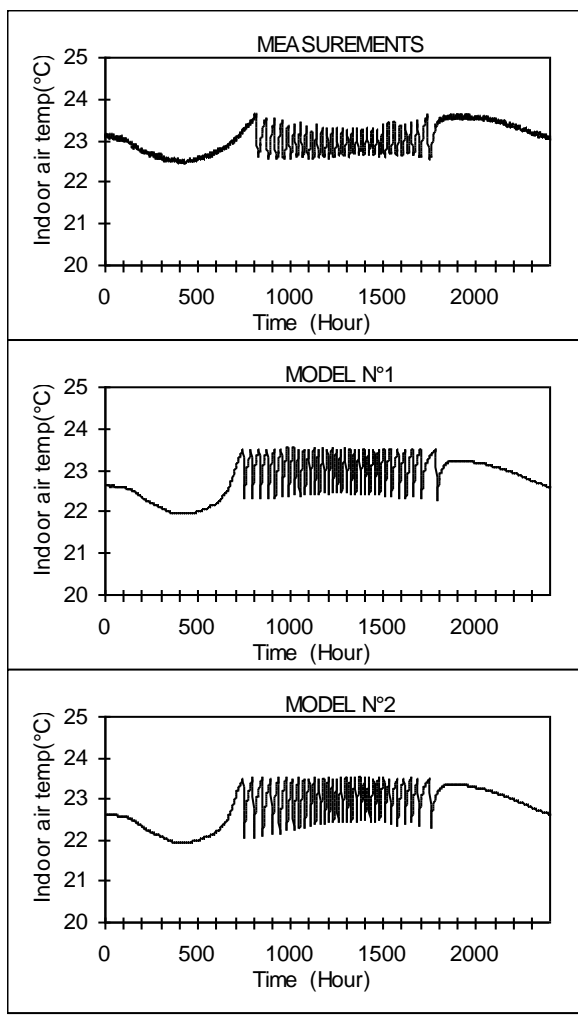
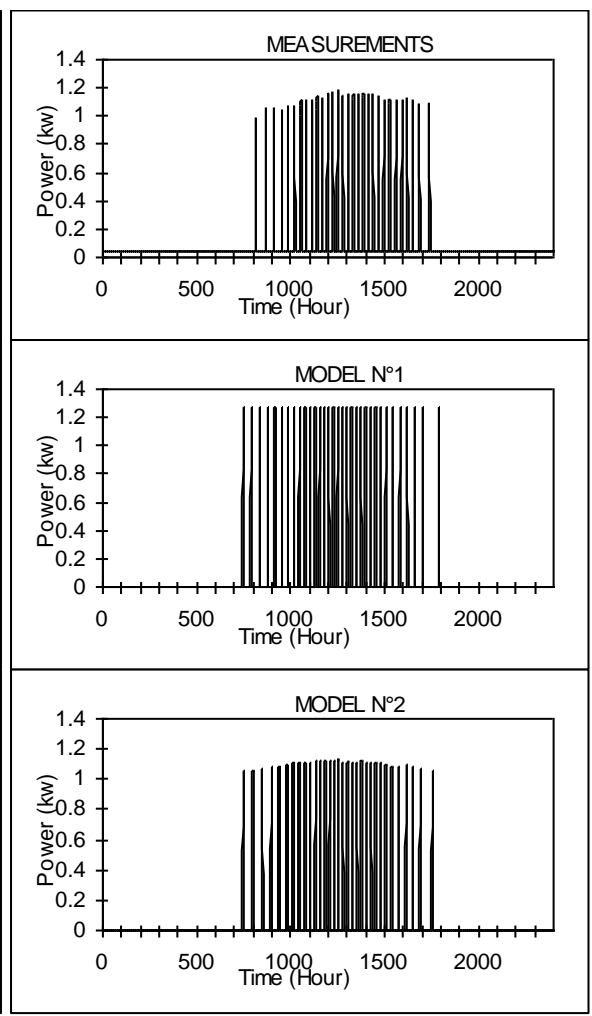

*Fig 7 : Indoor air temperature - time step =1 min*     *Fig 8 : Electric power - time step = 1 min*

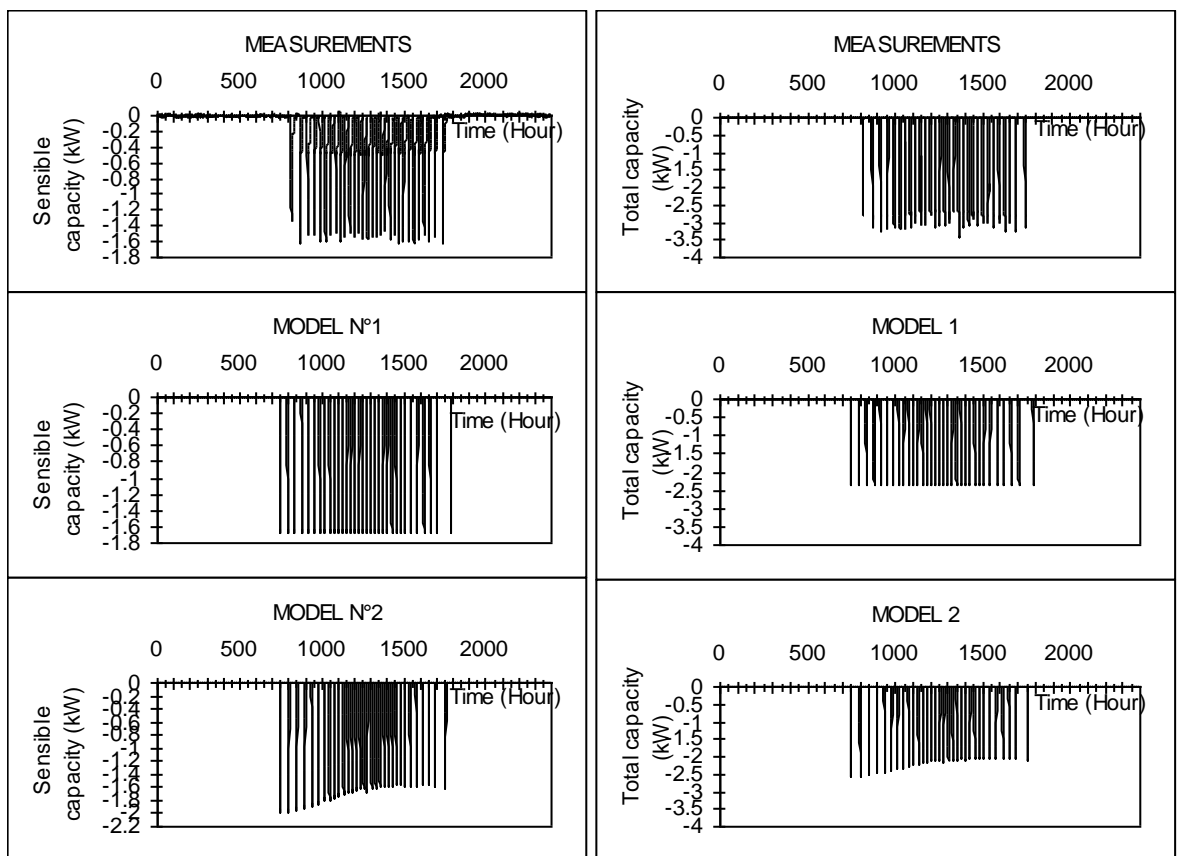

*Fig 9 : Sensible capacity - time step = 1 min*    *Fig 10 : Total cooling capacity - time step = 1 min*

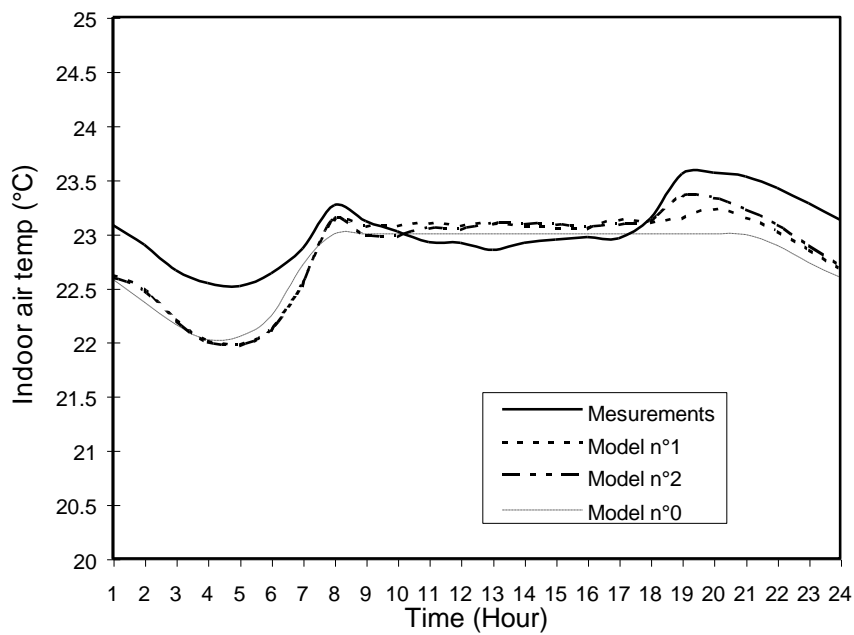

*Fig 11 : Indoor air temperature - Comparison of measurements and simulations - time step = 1 hour*

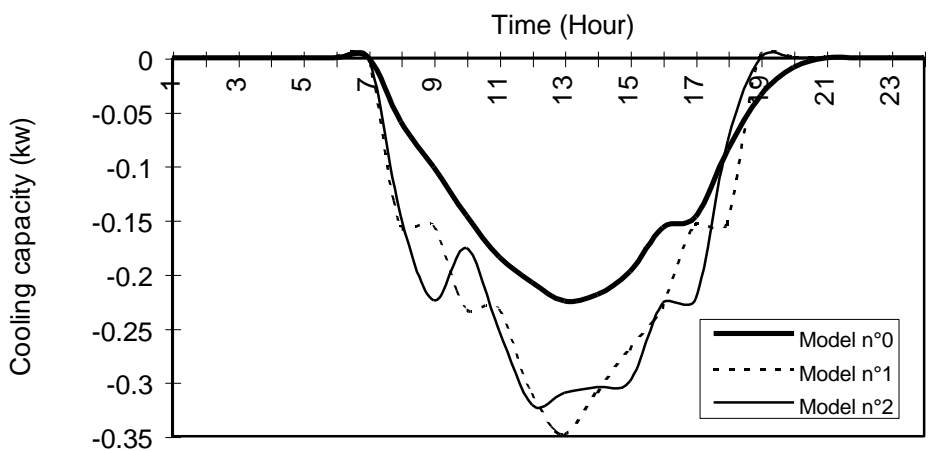

*Fig 12 : Total cooling capacity - Comparison of measurements and simulations - time step = 1 hour*

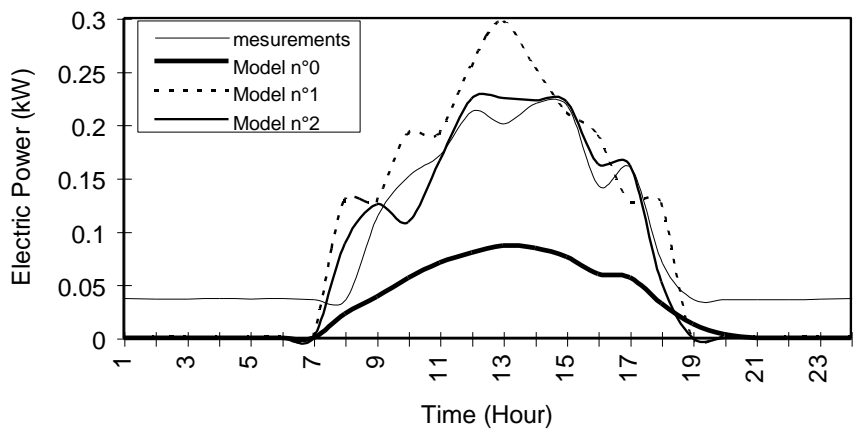

*Fig 13 : Electric power - Comparison of measurements and simulations - time step = 1 hour*

Visual inspection and comparison of the various graphs yields the following observations.
*Short time step (one minute):*

The comparison of the measured and simulated indoor air temperatures are very similar. The time of the first start-up of the split system is earlier for the models (approximatively around 7h30 a.m) than for the experiment (around 8h00). This is due to the 0.5°C gap between models and measurements during the passive period. Yet, the number of on/off cycling, the daily variations of the indoor air temperature and the last shut down (aroud 17h30) are well modellised.

One can see the same trends for the predicted power demand. The model n°2 seems to be more accurate than the model n°1 where the electric power is constant (it is supposed te be equal to the nominal power). Then power demand of Model n°1 is systematically bigger than the measured one.

The predictions of the sensible cooling capacity (see figure 9) show the correct general tendancies when compared to the measured values. On the contrary, there is a large gap between measurements values and the two models for the total cooling capacity (see figure 10). This gap is due to the fact that the time delay of one on-off cycle is shorter than the measured one. The modelled capacity has not time enough to reach the steady-state value.

*One hour time step :*

Indoor air temperature predictions exibit the same trends as the air temperature. This is quite obvious because these values are just the average of the one-minute-time-step values. The maximum gap is around 0.5°C.

The results given by figures 12 and 13 are quite interesting. The predictions of total cooling capacities for models 1 and 2 are larger than the one obtained for the model 0. These discrepancies can be attributed to the simulation of the controller which is taken into account in models n°1 and 2 (and supposed to be ideal in model n°0). The same trends are pointed out in figure 13 for the predictions of electric power demand. The model 2 seems to be the more accurate with an accuraccy of less than 10%. On the contrary, model n°0 underestimates the power demand in a very important way. We can also see the limits of hourly simulations to predict systems consumptions such as heat-pump systems with specific control strategies.

## 7. CONCLUSIONS

Three modelling levels of air-to-air residential heat-pumps have been defined and been integrated in a thermal building simulation software. The first model is just supposed to supply the demand of hourly cooling loads and electric demand with an ideal control loop (no delay between the sollicitations and the time response of the system) . The second model is taking into account the on/off cycle and the control processing. The third one is based on linear model determined from manufacturer data. Dynamic effects are taken into account in the last two models thanks to a single time constant model. The time step is reduced to one minute. The ouputs are the total, sensible and latent cooling capacities at each time step.
The first comparisons between measurements and simulations point out that a dynamic simulation with shorter time steps than an hour give better results on the estimation of electric consumption. This is mostly due to the kind of control process ( on/off controller ) of these air-conditionner systems. Netherveless, improvments have to be made concerning the estimation of the total cooling capacity. We have to face physical problem such as non homogeneity of the air inside the test cell and air infiltrations, what implies a time-gap between measures and models. Future research will focus on the introduction of a time delay due to theses physical phenomenon.